\documentstyle[amssymb,12pt,thmsa,sw20lart]{article}

\input{tcilatex}
\begin{document}

\title{Power-law behavior in signal scattering process in vertical granular chain
with light impurities}
\author{Aiguo Xu \\
Department of Physics and \\
Center for Strongly Correlated Materials Research, \\
Seoul National University, Seoul 151-742, Korea}
\maketitle

\begin{abstract}
We investigate the scattering process of impulse in vertical granular chain
with light impurities. When the perturbation is weak, the quantities
describing the reflection rate exhibit power law behavior with the impuity
depth. The exponent is nearly independent of $v_i$. When the perturbation is
very strong, the vertical chain shows similar behavior to that of the
horizontal chain, so the exponent is zero. Our numerical investigation
begins from the weak perturbation region and extend to the nonlinear region
and found a peak of the exponent. The difficulty in extending the numerical
investigation to a stronger perturbation region is analyzed.

{\bf PACS numbers:} 45.70-n, 43.25.+y,46.40.Cd

{\bf Key Words:} granular chain, scattering process, power-law behavior,
reflection rate
\end{abstract}

\section{Introduction}

In the last ten years physicists have been interested again in the study of
dry granular materials. Granular materials are ubiquitous around us. They
have many properties that are greatly different from those associated with
common solids, liquids and gases. These unique properties lead to many
applications. Many new ideas have emerged from this recent development.
These systems have been studied from various different viewpoints
experimentally, numerically, and theoretically\cite{1,2,3}. Nesterenko\cite
{4}noted that the propagation of a perturbation in loaded chain with
Hertzian contacts possesses soliton-like features. The soliton-like behavior
of the signal in two-dimensional granular beds was studied by Sinkovits and
Sen\cite{5,6}. The soliton-like properties of the signal in the chain of
iron balls were examined experimentally by Coste et al\cite{7}. Sen et al 
\cite{8} studied the soliton-like pulse in perturbed and driven Hertzian
chains and their possible applications in detecting buried impurities. For
the vertical granular chain under gravity with power-law-type contact force,
J. Hong et al \cite{9} showed that there are two kinds of propagating modes,
quasisolitary and oscillatory, depending on the strength of impulse. The
type of dispersion and the oscillation frequency, wavelength, and period
follow power laws in depth; J. Hong and Aiguo Xu\cite{10} found that the
signal speed follows a power law in depth and the exponent varies smoothly
from a saturant value to zero when the strength of the perturbation varies
from very weak to very strong.

To explore the prospects for the application of impulse for detecting
certain buried objects in granular beds, it is imperative that one must
acquire a detailed understanding of the its properties. Up to now many
problems on this topics are still openning. In present paper, we study the
scattering process of impulse when it encounters some other kind of grains
which we name impurities. The granular chain system can be described by the
following motion equations

\begin{eqnarray}
m_n\ddot{x}_n &=&V^{\prime }(x_{n-1},x_n)-V^{\prime }(x_n,x_{n+1})+m_ng, 
\nonumber \\
n &=&1,\ 2,\ \cdots ,\ N.  \label{motione1}
\end{eqnarray}
where $m_n$ is the mass of the $n$th grain, $V(x_n,x_{n+1})$ is the
inter-grain interaction between the $n$th and the$(n+1)$th grains. In this
paper we study the impulses in a special case with $V(x_{n-1},x_n)=[\eta
/(p+1)][\Delta _0-(x_n-x_{n-1})]^{p+1}$, where $\Delta _0$ is the diameter
of the spherical grain, $p$ the exponent of the power-law type contact
force, and $\eta $ the elastic constant of the grain under consideration.
Thus equation (\ref{motione1}) becomes

\begin{eqnarray}
m_n\ddot{x}_n &=&\eta \{[\Delta _0-(x_n-x_{n-1})]^p-[\Delta
_0-(x_{n+1}-x_n)]^p\}  \nonumber \\
+m_ng,n &=&1,\ 2,\ \cdots ,\ N.  \label{motione2}
\end{eqnarray}
For the Hertzian chain, i.e., the chain with Hertzian interaction between
neighboring granular spheres, $p=3/2$.

\section{Analytic treatment}

In our system most of the grains are same and the second kind of grains only
occur as impurities. Hence to analytically get some information of the
system, we first fix the mass of all the grains to a constant, i.e., $m_n=m$
for all $n$. Then we introduce a new variable $\psi _n$, denoting the
displacement of $n$th grain from equilibrium, defined by 
\begin{equation}
\psi _n=x_n-n\Delta _0+\sum_{l=1}^n\left( \frac{mgl}\eta \right) ^{1/p},
\label{vari}
\end{equation}
where the last term in the right-hand side is the sum of grain overlaps up
to $n$th contact and we set $x_0=\psi _0=0$. Eq. (\ref{motione2}) is
transformed into 
\begin{eqnarray}
m\ddot{\psi}_n &=&\eta \left[ \left( \frac{mgn}\eta \right) ^{1/p}+(\psi
_{n-1}-\psi _n)\right] ^p  \nonumber \\
&-&\eta \left[ \left\{ \frac{mg(n+1)}\eta \right\} ^{1/p}+(\psi _n-\psi
_{n+1})\right] ^p+mg  \label{origin}
\end{eqnarray}
using Eq. (\ref{vari}).

For the weakly nonlinear regime, the condition 
\begin{equation}
|\psi _{n-1}-\psi _n|\ll \left( \frac{mgn}\eta \right) ^{1/p}
\end{equation}
is valid and the expansion of Eq. (\ref{origin}) under this condition reads 
\begin{equation}
m\ddot{\psi}_n=-\mu _n(\psi _n-\psi _{n-1})+\mu _{n+1}(\psi _{n+1}-\psi _n)
\label{linear}
\end{equation}
where $\mu _n=mpg\left( \frac \eta {mg}\right) ^{1/p}n^{1-\frac 1p}$ is the
force constant of $n$th contact of the linear horizontal chain. We drop weak
nonlinear terms of the expansion in Eq. (\ref{linear}) to make the system
linear. Thus the vertical granular chain becomes a horizontal chain with
varying force constants in which the gravity effect is contained. Both left
and right side of Eq. (\ref{linear}) are linear in $\psi _n$. Therefore, the
scaling analysis tells us that the equation of motion (\ref{linear}) has
nothing to do with the initial impulse $v_i$. For this case we\cite{10} have
obtained the following power-law, 
\begin{equation}
v(h)\propto h^{-\frac 14\left( \frac 13+\frac 1p\right) }\text{.}
\end{equation}
The reducing of signal amplitude mainly due to the dispersion effect.

In our present studies two distances are important. The first one is the
depth of the impurities $d_{imp}$. The second one is the distance between
the impurity center and the grain where we get datas, $d$. Our main
objectives are to get some knowledge on the reflection rate and transmission
rate of the kinetic energy from the impurities. For fixed $d$, we expect the
reflection rate follows a power-law and the exponent is independent of $v_i$.

For the strongly nonlinear regime, the condition 
\begin{equation}
(\psi _{n-1}-\psi _n)\gg \left( \frac{mgn}\eta \right) ^{1/p}
\end{equation}
is valid and the expansion under this condition leads Eq. (\ref{origin}) to 
\begin{eqnarray}
m\ddot{\psi}_n &=&\eta \left[ (\psi _{n-1}-\psi _n)^p+pg_n(\psi _{n-1}-\psi
_n)^{p-1}\right]  \nonumber \\
&-&\eta \left[ (\psi _n-\psi _{n+1})^p+pg_{n+1}(\psi _n-\psi
_{n+1})^{p-1}\right]  \label{nonlinear}
\end{eqnarray}
where $g_n=\left( mgn/\eta \right) ^{1/p}$ denotes grain overlap at $n$th
contact. The gravity term can be neglected in the highly nonlinear regime,
since the gravity effect appears in the coefficient $g_n$. Different order
of $\psi _n$ in the left and right side of Eq. (\ref{nonlinear}) implies
that $v_i$-dependence must be appeared in the signal characteristics. Under
the limitation $v_i\rightarrow \infty $, Eq. (\ref{nonlinear}) recovers to

\begin{equation}
m\ddot{\psi}_n=\eta [(\psi _{n-1}-\psi _n)^p-(\psi _n-\psi _{n+1})^p]
\label{horizontal}
\end{equation}
That is to say, the vertical chain shows same behavior as that of the
horizontal chain. For the horizontal chain the signal speed is a constant,
so the exponent of the depth-dependence is $0$.

\section{Simulation results}

Our results are based on careful numerical integration of the coupled
equations of motion for granular chains with $N$ spheres, where $N$ varies
from $1000$ to $5000$ according to our need. The predictor-corrector method,
Runge-Kutta method, Richardson extrapolation and the Bulirsch-Stoer method%
\cite{11} are used. The program units are just the same as those in Refs.%
\cite{8} and \cite{9}. That is to say, the units of distance, mass, and time
are $10^{-5}m$, $2.36\times 10^{-5}kg$, and $1.0102\times 10^{-3}s$,
respectively. This kind of program units gives the gravitational
acceleration $g=1$. We set the grain diameter $100$ and $\eta =5657$.

Figure 1 shows four snapshots for a scattering process, where the impurity
depth is $d_{imp}=500$, which is described by the grain number; there are
five impurities around the depth $d_{imp}=500$; the mass of the impurity
grain is $m_{imp}=0.1$; the initial perturbation strength is $v_i=0.1$. Fig.
1 (a) shows a snapshot before the scattering process. Figs. 1 (b) and 1(c)
show the scattering process. Fig. 1 (d) shows a snapshot where the
scattering process nearly finished. We use ``1st'' and ``2nd'' to denote the
leading and second leading peaks. The leading and second leading peaks of
the incident impulse are pointed by an arrow, respectively, in Figs. 1 (a)
and 1(b). The leading and second leading peaks of the scattered impulse are
pointed by an arrow, respectively, in Figs. 1 (c) and 1 (d). It is clear
that the leading peak of the scattering impulse is upwards, which is a
characteristic property of the scattering process from light impurities\cite
{6}. We use five impurities because that the scattered impulse is too weak
if we use only one impurity. To understand the scattering process, we are
interested in the variation of the ratios $v_{ref}^{1st}/v_{inc}^{1st}$ , $%
v_{ref}^{2nd}/v_{inc}^{2nd}$ and $v_{ref}^{2nd}/v_{inc}^{1st}$with the depth
of the impurities $d_{imp}$, where $v_{inc}^{1st}$ and $v_{ref}^{1st}$ are
maximum values of the leading peaks of the incident and reflected impulses, $%
v_{inc}^{2nd}$ and $v_{ref}^{2nd}$ are maximum values of the second leading
peaks of the incident and reflected impulses. We obtain these datas from a
gain with depth $n=d_{imp}-d$. The three ratios describe the reflection rate
from different sides. During the scattering process the shape of the impulse
is not stable, so an appropriate distance $d$ is necessary. These ratios
describe the reflection rate of the kinetic energy in the vertical granular
chain. In order to get exact value for the leading and second leading peaks,
we use the smooth velocity-time curve of the measured gain, instead of the
rough snapshots.

Fig.2 shows the variations of $v_{ref}^{1st}/v_{inc}^{1st}$ , $%
v_{ref}^{2nd}/v_{inc}^{2nd}$ and $v_{ref}^{2nd}/v_{inc}^{1st}$ with the
impurity depth $d_{imp}$. Fig. 2 (a) is for $%
v_{ref}^{1st}/v_{inc}^{1st}(d_{imp})$ and $%
v_{ref}^{2nd}/v_{inc}^{2nd}(d_{imp})$, where $d=200$, $v_i=1$. The squares
are numerical results for $v_{ref}^{1st}/v_{inc}^{1st}(d_{imp})$ and the
circles are numerical results for $v_{ref}^{2nd}/v_{inc}^{2nd}(d_{imp})$.
The lines are corresponding fitting curves. Fig.2 (b) is for $%
v_{ref}^{2nd}/v_{inc}^{1st}(d_{imp})$, where $d=200$, $v_i=5$. The solid
squares are for numerical results and the line is the fitting curve. We use
the simplest power-law model, $v_{ref}/v_{inc}\propto d_{imp}^{-\beta }$ ,
to fit the numerical results. The fitting results suggest that the three
ratios follow power-laws when $d_{imp}$ is large enough. From Refs. \cite{9}
we know that when the initial perturbation is weak the propagating mode is
oscillatory, when the initial perturbation is strong the propagating mode is
quasisolitary. It is interesting to check if the power-law model works for
both the weak and strong perturbations. If it works, we have the following
expected result: When $v_i$ is large enough, the behavior of the vertical
chain is similar to that of the horizontal chain; hence there is no
power-law. That means $\beta \rightarrow 0$ as $v_i\rightarrow \infty $.

Fig.3 shows the variation of the exponent for $v_{ref}^{2nd}/v_{inc}^{2nd}$
with the strength of initial perturbation, where the line with squares is
for the case of $d=100$, the line with circles is for the case of $d=200$.
The two curves show similar behaviors. It is clear that the exponent is
independent of $v_i$ when $v_i$ is very small. For the case of $d=100$, $%
\beta =0.19\pm 0.01$. For the case of $d=200$, $\beta =0.23\pm 0.01$. In
fact when $v_i$ is too small we can hardly get helpful information from the
simulation datas due to numerical errors. The curve for $d=100$ has a peak
around $\tilde{v}_i=5$. The curve for $d=200$ has a peak around $\tilde{v}%
_i=10$. When $v_i>\tilde{v}$ , $\beta $ decreases quickly with the
increasing of $v_i$. We tried to add more points to the figure, but when $%
v_i $ is larger than $20$ we can hardly measure the second leading peak from
the incident impulse. We can know the reason from Fig. 4.

Fig. 4 shows a $v_n(t)$ relation, where $d_{imp}=300$, $d=200$, $n=100$, $%
v_i=15$, and $m_{imp}=0.1$. The impulse near $t=2$ is the incident one. The
impulse within $9<t<10$ is the reflected one (Only part of the reflected
impulse is plotted). It is clear that for a strong perturbation the
propagating impulse is quasisolitary and it has no evident the second
leading peak; at the same time, the leading peak of the reflected impulse is
very small and there exist some oscillations before the leading peak occurs.

When the perturbation is weak, the propagation of the wave is oscillatory;
hence the two ratios $v_{ref}^{1st}/v_{inc}^{1st}$ and $%
v_{ref}^{2nd}/v_{inc}^{2nd}$ can describe well the reflection rate of the
kinetic energy from the impurities. But with the increasing of $v_i$, the
propagation of the wave is more and more quasisolitary and the ratio $%
v_{ref}^{2nd}/v_{inc}^{1st}$ describes the reflection rate better. Fig.5
shows the variation of the exponent $\beta ^{\prime }$ for $%
v_{ref}^{2nd}/v_{inc}^{1st}(d_{imp})$ with $v_i$, where $d=200$. It shows
similar behavior to the exponent $\beta $ for $%
v_{ref}^{2nd}/v_{inc}^{2nd}(d_{imp})$. When $v_i$ is very small, $\beta
^{\prime }\thickapprox 0.2$.

Although we have found the power-law behavior for $%
v_{ref}^{1st}/v_{inc}^{1st}(d_{imp})$ for not very large $v_i$ (see Fig. 2
for an example), it is hard for us to get a figure which is similar to Fig.3
and Fig.5. The reasons will be given in the following part.

For a fixed distance $d$, when the depth of impurities $d_{imp}$ is large,
we find the curve $v_{ref}^{1st}/v_{inc}^{1st}(d_{imp})$ oscillates greatly
with $d_{imp}$. Fig. 6 shows two examples, where $d=200$; $v_i=6$ in (a) and 
$v_i=4$ in (b). We can understand the reason for this kind of behaviors
through Fig. 7. Fig.7 shows several $v_n(t)$ curves from which we can study
the scattering process of the impulse, where $n=d_{imp}-100$, $v_i=0.1$, $%
m_{imp}=0.1$; $d_{imp}=600$ in (a), $d_{imp}=800$ in (b), $d_{imp}=1000$ in
(c), and $d_{imp}=1100$ in (d), respectively. For the reflected impulse,
only the leading peak and the second peak are shown. For a fixed value of $d$%
, when the depth $d_{imp}$ is small, the reflected impulse and the incident
impulse are separated, i.e. there is a time during which the measured grain
is static. For these cases, we can get exact values for $v_{inc}^{1st}$, $%
v_{ref}^{1st}$, $v_{inc}^{2nd}$, and $v_{ref}^{2nd}$. When the depth $%
d_{imp} $ is large enough, the tail of the incident impulse and the
reflected impulse are connected. For these cases, we can hardly get exact
values for $v_{ref}^{1st}$ and $v_{ref}^{2nd}$. From Fig.7 we also
understand the fact these phenomena first influence the measurement of $%
v_{ref}^{1st}$ , then $v_{ref}^{2nd}$.

We name the increasing process of $d_{imp}$ before a power-law occurs a
transient process. The simulation results show that the ratio $%
v_{ref}^{1st}/v_{inc}^{1st}(d_{imp})$ needs a much longer transient process
than $v_{ref}^{2nd}/v_{inc}^{2nd}(d_{imp})$. The initial impulse is
stronger, we need a longer transient process. The curves for $%
v_{ref}^{2nd}/v_{inc}^{2nd}(d_{imp})$ decreases with the increasing of $%
d_{imp}$, while the curves for $v_{ref}^{1st}/v_{inc}^{1st}(d_{imp})$ first
increases, then decreases. The initial impulse is stronger, the curves
decrease more slowly. We also notice that there are more fitting errors when 
$v_i$ is large.

\section{Conclusion}

In present paper we study the scattering process of impulse in vertical
granular chain with light impurities. Most of the studies resort on
numerical simulations and the analytic treatment helps to understand the
behaviors. When the perturbation is weak, the quantities describing the
reflection rate exhibit power law behavior with the impuity depth. The
exponent is independent of $v_i$. When the perturbation is very strong, the
vertical chain shows similar behavior to that of the horizontal chain, so
the exponent is zero. Our numerical investigation begins from the weak
perturbation region and extend to the nonlinear region and found a peak of
the exponent. The difficulty in extending the numerical investigation to a
stronger perturbation region is analyzed. More work needs to be done to
completely understand the scattering process of the signal in vertical
granular chain.

{\bf Acknowledgments}

The author acknowledges Prof. Jongbae Hong for helpful suggestions and
discussions. This work was supported by Korea Research Foundation Grant
(KRF-2000-DP0106). The author also wishes to acknowledge the support by the
BK21 project of the Ministry of Education, Korea.

\newpage

{\bf Figure caption:}

{\bf Figure 1: }Snapshots for a scattering process, where the impurity depth
is $d_{imp}=500$, which is described by the grain number; there are five
impurities around the depth $d_{imp}=500$; the mass of the impurity grain is 
$m_{imp}=0.1$; the initial perturbation strength is $v_i=0.1$.

{\bf Figure 2: }Variations of $v_{ref}^{1st}/v_{inc}^{1st}$ , $%
v_{ref}^{2nd}/v_{inc}^{2nd}$ and $v_{ref}^{2nd}/v_{inc}^{1st}$ with the
depth of the impurities $d_{imp}$. Fig.2 (a) is for $%
v_{ref}^{1st}/v_{inc}^{1st}(d_{imp})$ and $%
v_{ref}^{2nd}/v_{inc}^{2nd}(d_{imp})$ where $d=200$, $v_i=1$. The squares
are numerical results for $v_{ref}^{1st}/v_{inc}^{1st}(d_{imp})$ and the
circles are numerical results for $v_{ref}^{2nd}/v_{inc}^{2nd}(d_{imp})$.
The lines are corresponding fitting curves. Fig.2 (b) is for $%
v_{ref}^{2nd}/v_{inc}^{1st}(d_{imp})$ where $d=200$, $v_i=5$. The solid
squares are for numerical results and the line is the fitting curve.

{\bf Figure 3:} Variation of the exponent for $v_{ref}^{2nd}/v_{inc}^{2nd}$
with the strength of initial perturbation, where the line with squares is
for the case of $d=100$, the line with circle is for the case of $d=200$.

{\bf Figure 4:} The $v_n(t)$ relation, where $d_{imp}=300$, $d=200$, $n=100$%
, $v_i=15$, and $m_{imp}=0.1$. The impulse near $t=2$ is the incident one.
The impulse within $9<t<10$ is the reflected one (Only part of the reflected
impulse is plotted). The second leading peak of the incident impulse and the
leading peak of the reflected impulse are very small. Some oscillations
occur behind the main incident impulse.

{\bf Figure 5: }Variation of the exponent $\beta ^{\prime }$ for $%
v_{ref}^{2nd}/v_{inc}^{1st}(d_{imp})$ with $v_i$, where $d=200$. It shows
similar behavior to the exponent $\beta $ for $%
v_{ref}^{2nd}/v_{inc}^{2nd}(d_{imp})$. When $v_i$ is very small, $\beta
^{\prime }\thickapprox 0.2$.

{\bf Figure 6: }Reflection rate $v_{ref}^{1st}/v_{inc}^{1st}$ versus $%
d_{imp} $, where $d=200$; $v_i=6$ in (a) and $v_i=4$ in (b).

{\bf Figure 7: }Velocity{\bf \ }$v_n(t)$ curves, where $n=d_{imp}-100$, $%
v_i=0.1$, $m_{imp}=0.1$; $d_{imp}=600$ in (a), $d_{imp}=800$ in (b), $%
d_{imp}=1000$ in (c), and $d_{imp}=1100$ in (d), respectively.

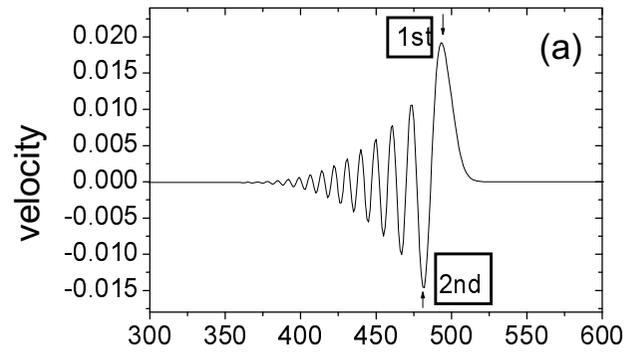
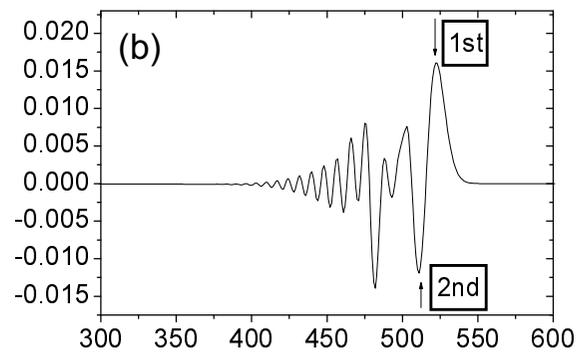
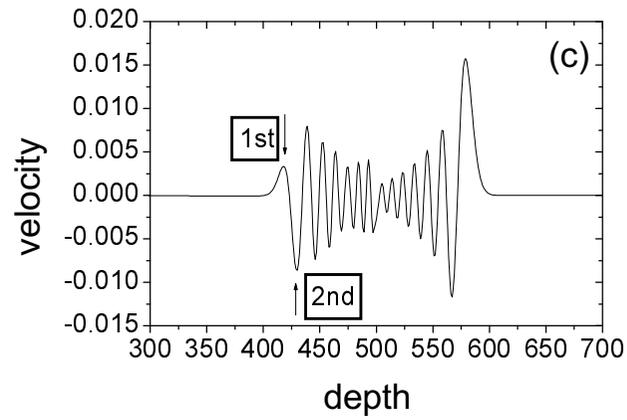
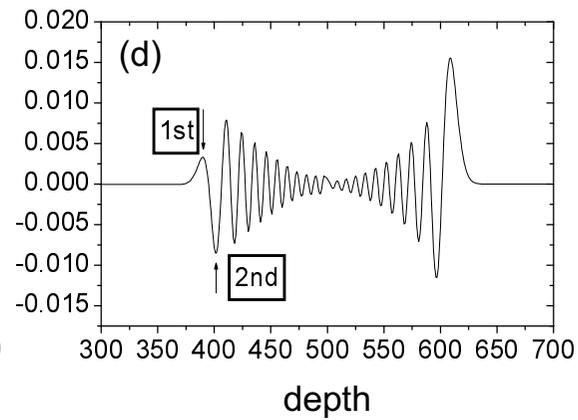
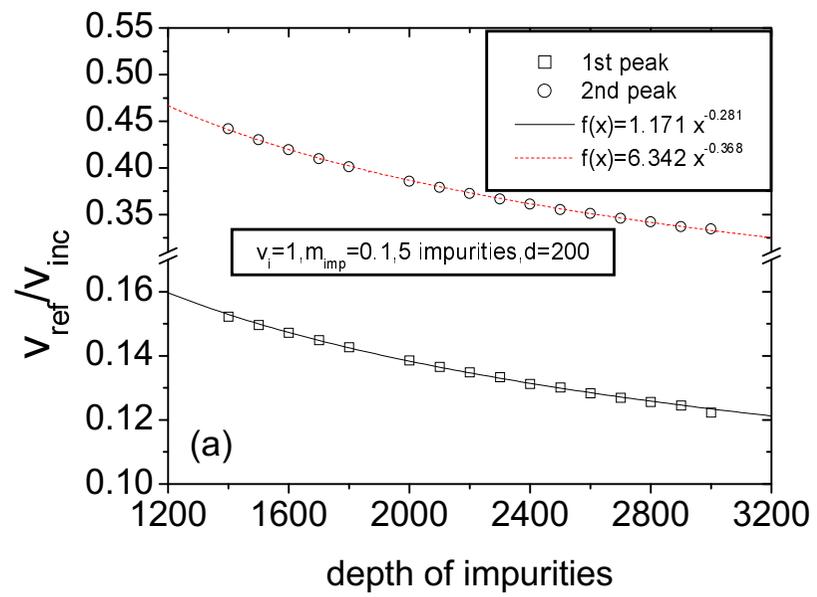
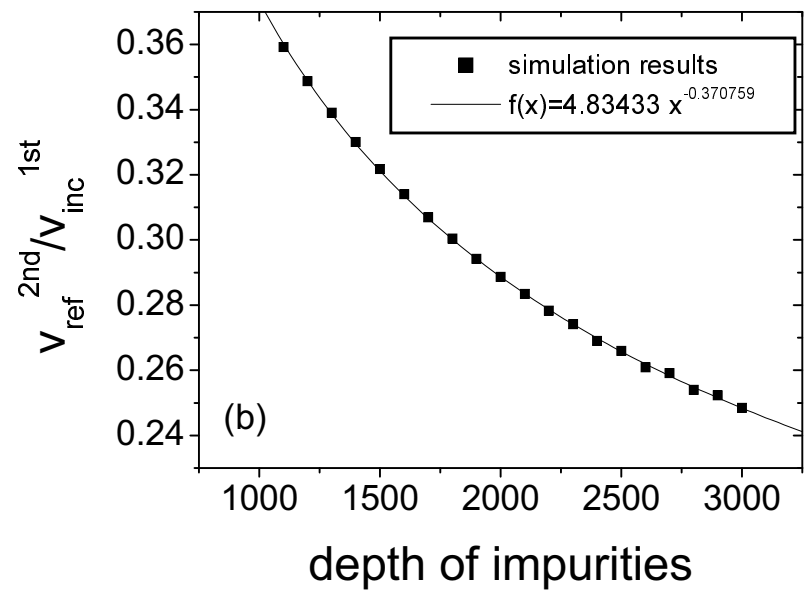
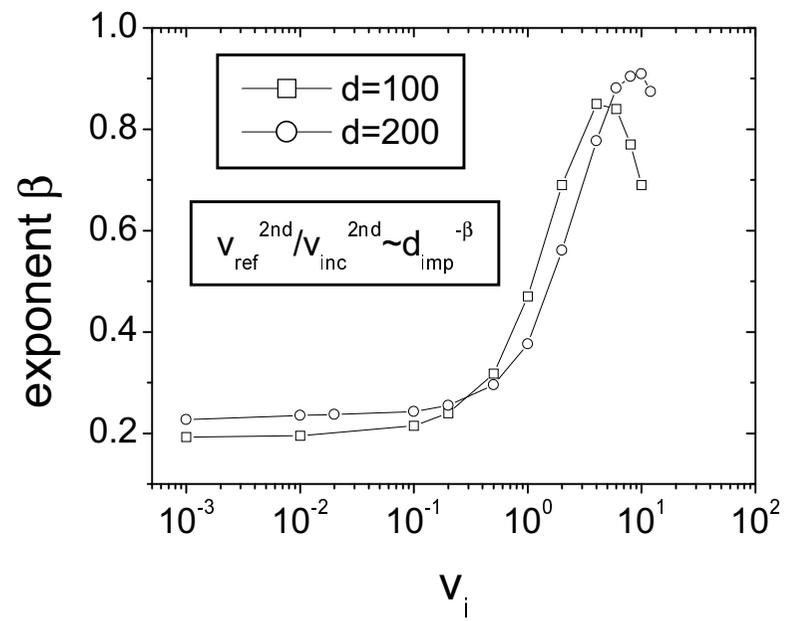
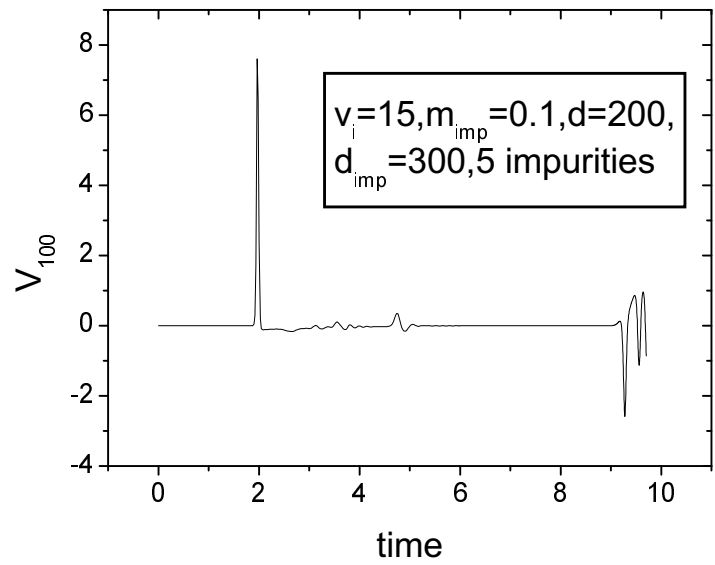
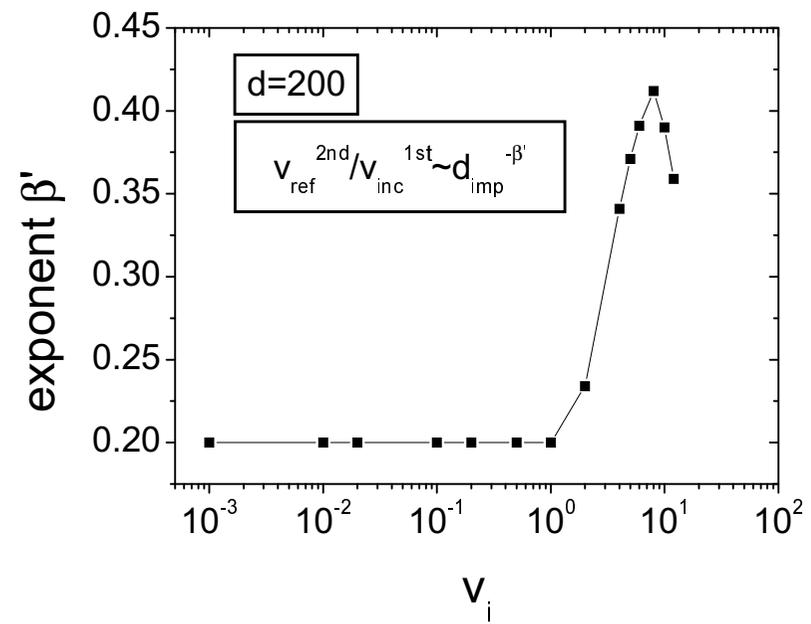
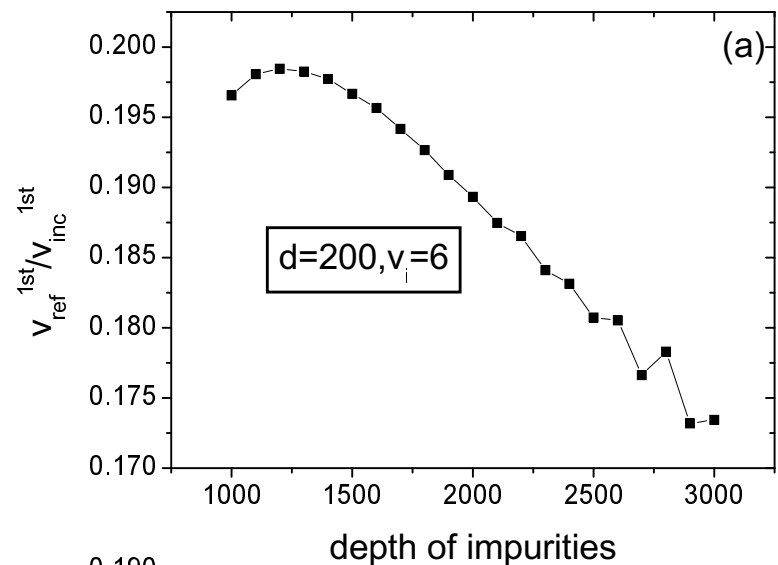
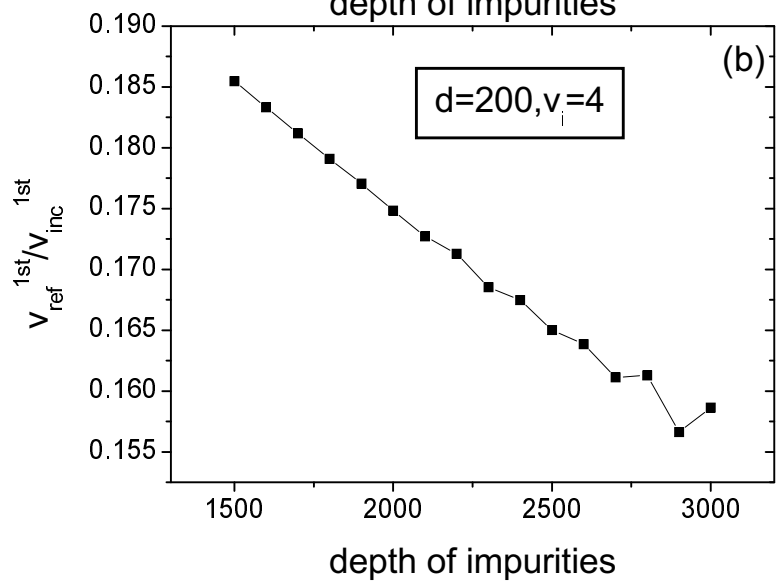
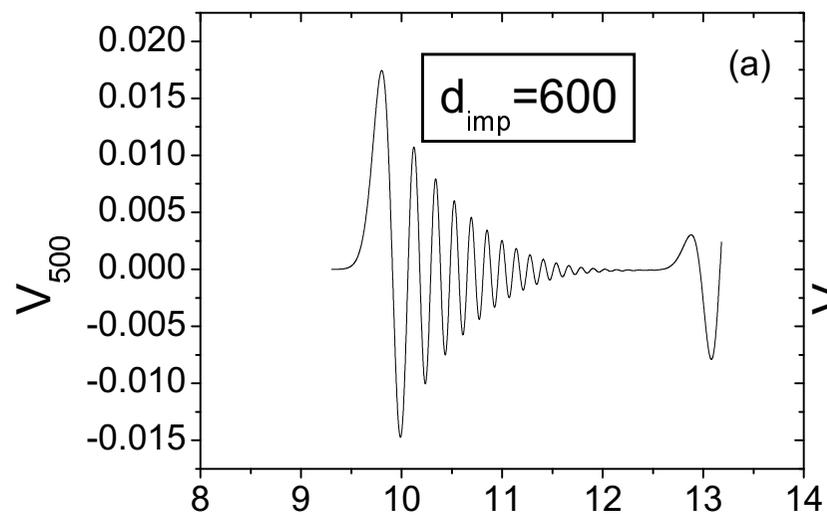
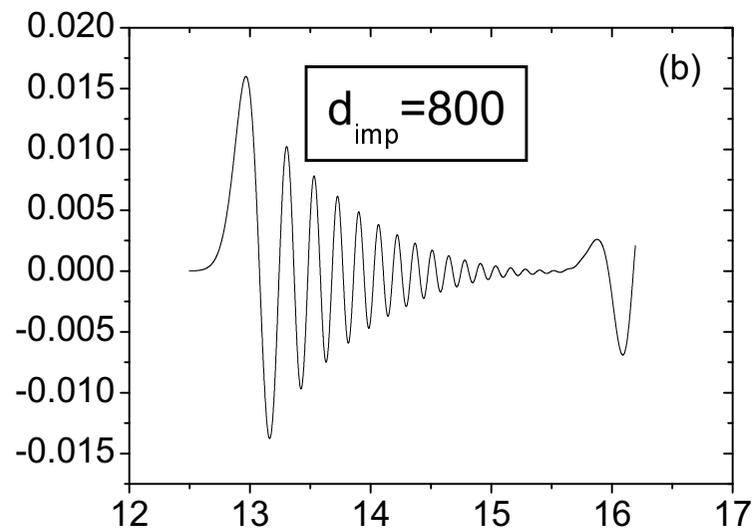
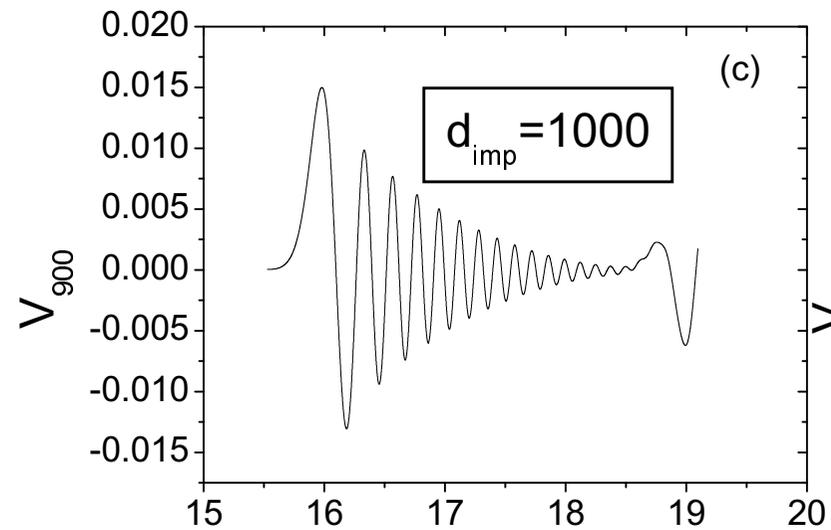
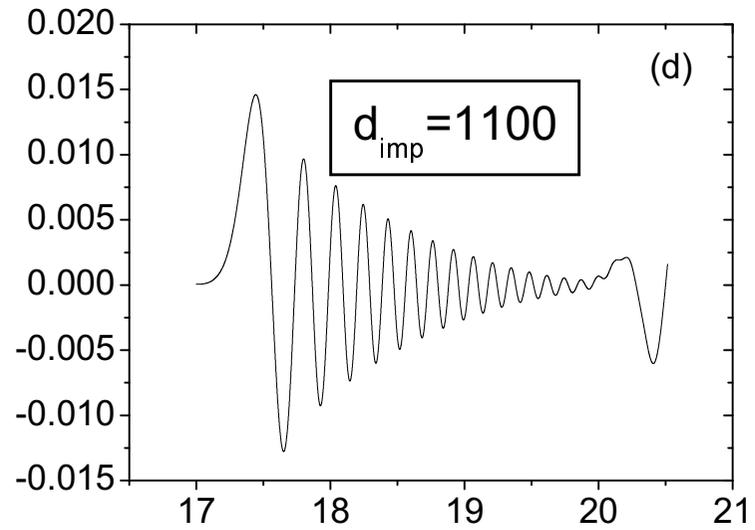
\end{document}